\documentclass{aa}

\usepackage[varg]{txfonts}
\usepackage{longtable}
\usepackage{bm}

\newcommand\kms{km~s$^{-1}$}
\newcommand\msun{$M_\odot$}
\newcommand\mhi{$M_{\mathrm{HI}}$}
\newcommand\mstar{$M_{*}$}
\newcommand\lsun{$L_\odot$}

\def\be{\begin{equation}}
\def\ee{\end{equation}}
\def\a40{$\alpha$.40}
\def\arcmin{$^{\prime}$}
\def\arcsec{$^{\prime\prime}$}
\def\dg{$^{\circ}$}

\newcommand{\hi}{H\,{\sc i}}
\def\mdyn{$M_{dyn}$}

\defcitealias{2013ApJ...768...77A}{A13}
\defcitealias{2015ApJ...800L..15B}{B15}
\defcitealias{2015ApJ...806...95S}{S15}

\usepackage{natbib,twoopt}
\usepackage[breaklinks=true]{hyperref} %% to avoid \citeads line fills
\bibpunct{(}{)}{;}{a}{}{,} %% natbib format for A&A and ApJ
\makeatletter
\newcommandtwoopt{\citeads}[3][][]{\href{http://adsabs.harvard.edu/abs/#3}%
{\def\hyper@linkstart##1##2{}%
\let\hyper@linkend\@empty\citealp[#1][#2]{#3}}}
\newcommandtwoopt{\citepads}[3][][]{\href{http://adsabs.harvard.edu/abs/#3}%
{\def\hyper@linkstart##1##2{}%
\let\hyper@linkend\@empty\citep[#1][#2]{#3}}}
\newcommandtwoopt{\citetads}[3][][]{\href{http://adsabs.harvard.edu/abs/#3}%
{\def\hyper@linkstart##1##2{}%
\let\hyper@linkend\@empty\citet[#1][#2]{#3}}}
\newcommandtwoopt{\citeyearads}[3][][]%
{\href{http://adsabs.harvard.edu/abs/#3}
{\def\hyper@linkstart##1##2{}%
\let\hyper@linkend\@empty\citeyear[#1][#2]{#3}}}
\makeatother

\begin{document}

\title{AGC 226067: A possible interacting low-mass system}
\author{
E.\ A.\ K.\ Adams
              \inst{1}            
                 \and
                  J.\ M.\ Cannon
               \inst{2}
		\and
	K.\  L.\ Rhode
               \inst{3}
               \and
               W.\ F.\ Janesh
               \inst{3} 
               \and
               S. Janowiecki
               \inst{3}
               \and
               L. Leisman
                \inst{4}
                \and
               R. Giovanelli
               \inst{4}
               \and
               M. P. Haynes
               \inst{4}
                \and
                T. A. Oosterloo
                \inst{1}
                \and
                J. J. Salzer
                \inst{3}
                \and                
		T. Zaidi
		\inst{2}
}
\institute{
ASTRON, Netherlands Institute for Radio Astronomy, Postbus 2, 7900 AA Dwingeloo, The Netherlands
 		\email{adams@astron.nl}
 		\and
		Department of Physics and Astronomy, Macalaster College, 1600 Grand Avenue, Saint Paul, MN 55105, USA
		\and
		Department of Astronomy, Indiana University, 727 East Third Street, Bloomington, IN 47405, USA		
		\and
		Center for Radiophysics and Space Research, Space Sciences Building, Cornell University, Ithaca, NY 14853, USA
		}

%%%%%%%%%%%%%%%%%%%%%%%%%%%%%%%%%

\abstract
{
We present Arecibo, GBT, VLA and WIYN/pODI observations of the ALFALFA source AGC 226067.
Originally identified as an ultra-compact high velocity cloud
and candidate Local Group galaxy, AGC 226067 is 
spatially and kinematically coincident with the Virgo cluster,
and the identification by multiple groups of an optical counterpart with
no resolved stars supports the interpretation that this systems lies at the Virgo distance ($D=17 $ Mpc).
The combined observations reveal that the system consists of multiple components: a central
\hi\ source associated with the optical counterpart (AGC 226067), a smaller \hi-only component 
(AGC 229490), a second optical component (AGC 229491), and extended low surface brightness \hi. 
Only $\sim$1/4 of the single-dish \hi\ emission is associated
with AGC 226067; 
as a result, we find \mhi/$L_{g} \sim6$ \msun/\lsun\,
which is lower than
previous work.
At $D=17$ Mpc, AGC 226067 has
an \hi\ mass of  $1.5 \times 10^7$ \msun\ and $L_{g} = 2.4 \times 10^6$ \lsun,
AGC 229490 (the \hi-only component) has  $M_{HI} = 3.6 \times 10^6$ \msun, and AGC 229491 (the second optical component) has $L_{g} = 3.6 \times 10^5\, L_{\odot}$.
The nature of this system of three sources is uncertain: AGC 226067 and AGC 229490 
may be connected by an \hi\ bridge, and AGC 229490 and AGC 229491 are separated by only 0.5\arcmin.
The current data do not resolve the \hi\ in AGC 229490 and its origin is unclear.
We discuss possible scenarios for this system of objects:
an interacting system of dwarf galaxies, accretion of material onto AGC 226067, or stripping
of material from AGC 226067. 
}


\maketitle

%%%%%%%%%%%%%%%%%%%%%%%%%%%%%%%

\section{Introduction}\label{sec:intro}

The ALFALFA \hi\ survey offers a census of neutral hydrogen (\hi) in the nearby Universe
\citepads{2005AJ....130.2598G}.
Due to its sensitivity and footprint area, ALFALFA has been able to detect
rare but extremely interesting objects, including \hi\ sources
that probe the extreme limits of systems that are able to form stars. 
These sources can be coarsely separated into three types:
the candidate gas-rich but (nearly) starless Local Group galaxies,
called ``ultra-compact high velocity clouds" or ``UCHVCs"  
\citepads[][hereafter A13]{2010ApJ...708L..22G,2013ApJ...768...77A};
the low-mass, gas-rich ``SHIELD" dwarf galaxies \citepads{2011ApJ...739L..22C};
the clearly extragalactic \hi\ detections with no discernible optical counterpart in extant
optical surveys, called ``Almost-Dark'' sources 
\citepads[][Leisman et al., in prep.]{2015ApJ...801...96J, 2015AJ....149...72C}.

AGC 226067 is an  ALFALFA source that highlights 
the  extreme objects contained within these samples.
It was included in the UCHVC
catalogs of \citetads{2010ApJ...708L..22G} and \citetalias{2013ApJ...768...77A} (as HVC274.68+74.70-123) due to its
low recessional velocity ($cz_{\odot} = -128$ \kms), small angular extent (5\arcmin\ $\times$ 4\arcmin)
and isolation from other high-velocity clouds. 
However, AGC 226067 is projected
onto the Virgo Cluster along with at least 14 galaxies located within 3\dg\ and 100 \kms. Two of these systems have Tully-Fisher distances consistent
with the distance to Virgo: NGC 4396  \citepads[][$D=15.8$ Mpc]{1999MNRAS.304..595G} and IC 3311 \citepads[][$D = 21.1$ Mpc]{2010Ap&SS.325..163P}. 
The detection by \citetads[][hereafter B15]{2015ApJ...800L..15B} and 
\citetads[][hereafter S15]{2015ApJ...806...95S} of an optical counterpart
to AGC 226067
without resolved stars in ground-based imaging implies $D\gtrsim 3$ Mpc, 
negating its identification as an UCHVC and 
making the association with the Virgo Cluster likely.  
We assume throughout this work that AGC 226067 is located in the Virgo Cluster at $D=17$ Mpc.
If it is located closer, it is a lower mass system.

In this paper we present global single-dish \hi\ spectra, resolved \hi\ imaging with the {\it Karl G. Jansky} Very Large Array (VLA), and deep optical imaging with WIYN/pODI of AGC 226067.
Our observations definitively associate the \hi\ with the optical counterpart as a low mass dwarf irregular, analogous to the SHIELD
galaxies.
Importantly,
these data reveal that AGC 226067 is an extended interacting system
with multiple components, both \hi\ and optical.

%%%%%%%%%%%%%%%%%%%%%%%%%%%%%%%%%%%%%%%%%
\section{Data}\label{sec:data}

\subsection{Single-dish \hi\ data}\label{sec:sd}
AGC 226067 was originally detected in the Arecibo Legacy Fast ALFA (ALFALFA) \hi\ line survey
and is included in the catalogs of \citetads{2007AJ....133.2569G} and \citetads{2011AJ....142..170H}; full
details of source extraction 
 are given in those works.
Due to the lack of an optical counterpart and the low recessional velocity, this source
was also included in the UCHVC catalogs of \citet{2010ApJ...708L..22G} and \citetalias{2013ApJ...768...77A}.
In the work of \citetalias{2013ApJ...768...77A}, the sources were independently re-measured to ensure that the extended, diffuse
emission common to the UCHVCs was properly included, resulting in a flux density
that was 15\% higher than given by \citetads{2011AJ....142..170H}.

\begin{figure}
\centering
\includegraphics[width=\linewidth,trim=1cm 3cm 1cm 3cm,clip=true]{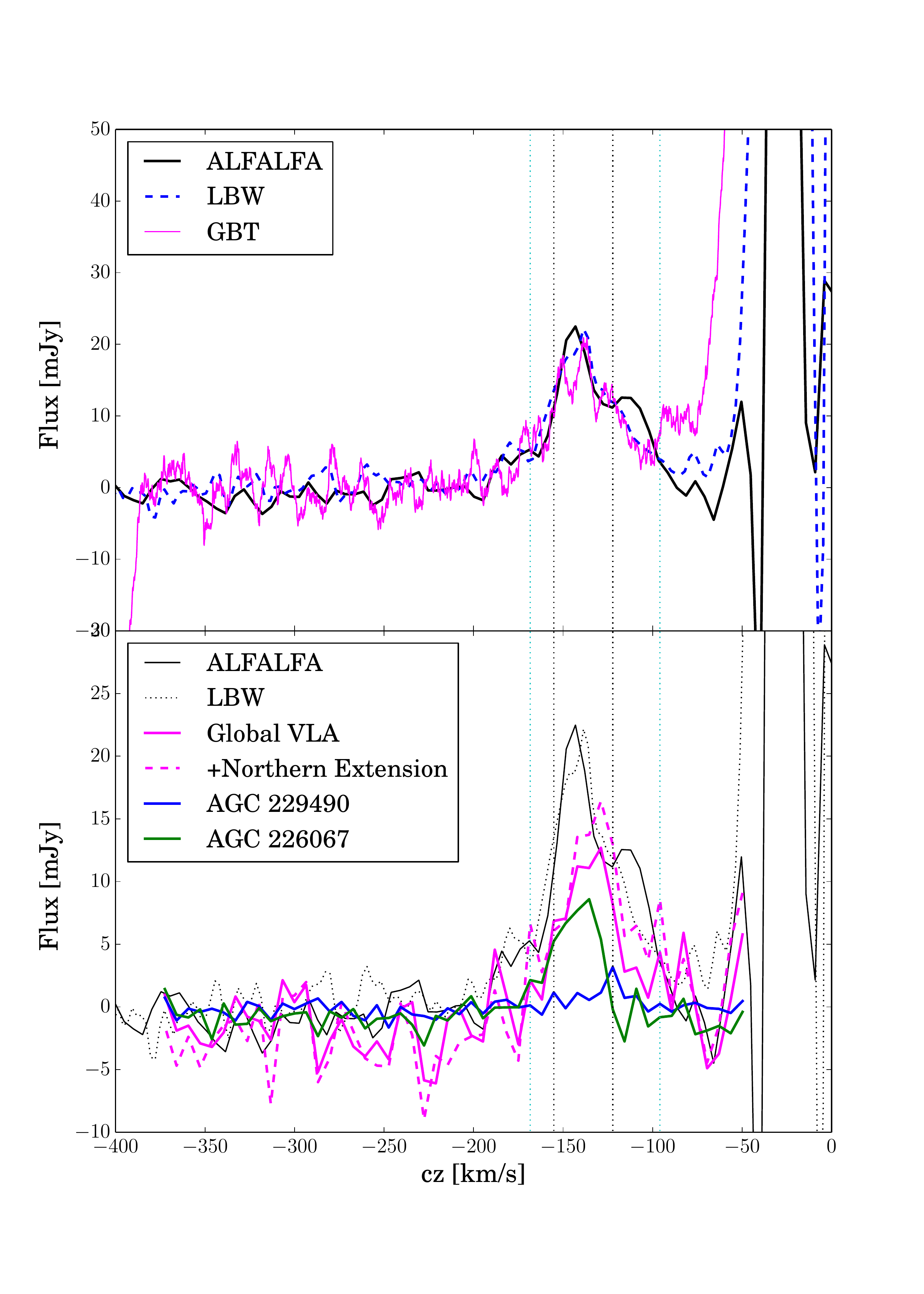}
\caption{{\it Top:} The single-dish spectra for AGC 226067.
Both the ALFALFA and LBW spectrum are Hanning smoothed to a velocity resolution of 10 \kms, and
the GBT spectrum is boxcar smoothed to a resolution of 5 \kms.
%The VLA spectrum from an aperture matched to the AO beam (4\arcmin) is also shown.
Dotted lines indicate the two velocity ranges used for constructing the VLA total \hi\ intensity maps.
{\it Bottom:} The global VLA spectrum, including with the potential northern extension, along with the isolated spectra for AGC 226067 and AGC 229490.
The ALFALFA and LBW spectra are included for reference.}
\label{fig:spec}
\end{figure}

AGC 226067 was also observed with both the L-band wide receiver (LBW) at Arecibo Observatory (AO; 22 March 2013, program A2752) and the {\it Robert C. Byrd} Green Bank Telescope (GBT; 23 October 2012, program GBT/12B-059) as part of programs aimed at confirming UCHVCs.
The LBW observation was  a 3 minute ON-OFF position switching pair, and the GBT
observation was a 15 minute track in frequency-switching mode.
Both spectra were calibrated with observatory-provided routines and the
baseline-subtracted spectra are shown in the upper panel of Figure \ref{fig:spec},
along with the ALFALFA spectrum from \citetalias{2013ApJ...768...77A}.

Table \ref{tab:globalhi} reports the global \hi\ properties from the single-dish spectra.
The LBW spectrum matches the ALFALFA spectrum well but lacks the bump in emission in the red-wing of the
profile.
Consequently, the measured velocity width and flux density are slightly lower but still consistent with the 
ALFALFA data.
The ALFALFA source size is more extended than the AO beam, so the missing emission could 
be 
outside the LBW pointing. 
However, the GBT beam encompasses the full source extent of AGC 226067.
While the GBT spectrum is much noisier, it 
is a good match to the LBW spectrum, indicating that the 
lack of red-wing emission may be real and the majority of emission is located within the AO beam. 
Additionally, all the spectra show low amplitude blue-wing emission, indicating that this may be true emission.

\begin{table}
\footnotesize
\caption{Global \hi\ Properties}
\label{tab:globalhi}
\centering
\begin{tabular}{l l l l}
\hline
\hline
 Data &  $S_{HI}$ & $cz_{\odot}$& $W_{50}$ \\
 &  Jy \kms & \kms & \kms   \\
\hline
 ALFALFA \tablefootmark{a} & 0.92 $\pm$ 0.1 & -128 $\pm$ 4&54 $\pm$ 13 \\
 LBW & 0.80 $\pm$ 0.15 & -139 $\pm$ 1 & 48 $\pm$ 6  \\
 GBT &  0.77 $\pm$ 0.18  & -139 $\pm$ 2   &  48 $\pm$ 4\\
 VLA & 0.47 $\pm$ 0.20 & -135 $\pm$ 3 & 34 $\pm$ 3 \\
 $\,$+ Northern extension & 0.68 $\pm$ 0.20 & -133 $\pm$ 3 & 35 $\pm$ 6\\
% 0.50  %\tablefootmark{b}   
 % (0.75)\tablefootmark{b} & -139 $\pm$ 3 & 40 $\pm$ 5 \\
%Extension (VLA) & -- & 0.25 & \\
\hline
\end{tabular}
\tablefoot{
\tablefoottext{a}{From \citetalias{2013ApJ...768...77A}}
%\tablefoottext{b}{Central VLA emission only}
%\tablefoottext{b}{Including the northern extension}
%\tablefoottext{a}{Unresolved by VLA beam}
}
\end{table}

\subsection{VLA \hi\ data}\label{sec:vla}
AGC 226067 was observed in July 2014 with the VLA in D-configuration for 
 3 one-hour blocks, corresponding to 1.7 hours of on-source integration time.
The WIDAR correlator was configured to provide a 4 MHz bandwidth subdivided into 1024 channels,
corresponding to a native velocity resolution of 0.82 \kms\ ch$^{-1}$.
Standard calibration of the visibility data was done in  AIPS, including bandpass, flux
and complex gain calibration. Line-free channels were used to remove the continuum
emission.

\begin{figure*}
\centering
\includegraphics[width=.9\linewidth,trim=0cm 0cm 0cm 0cm,clip=true]{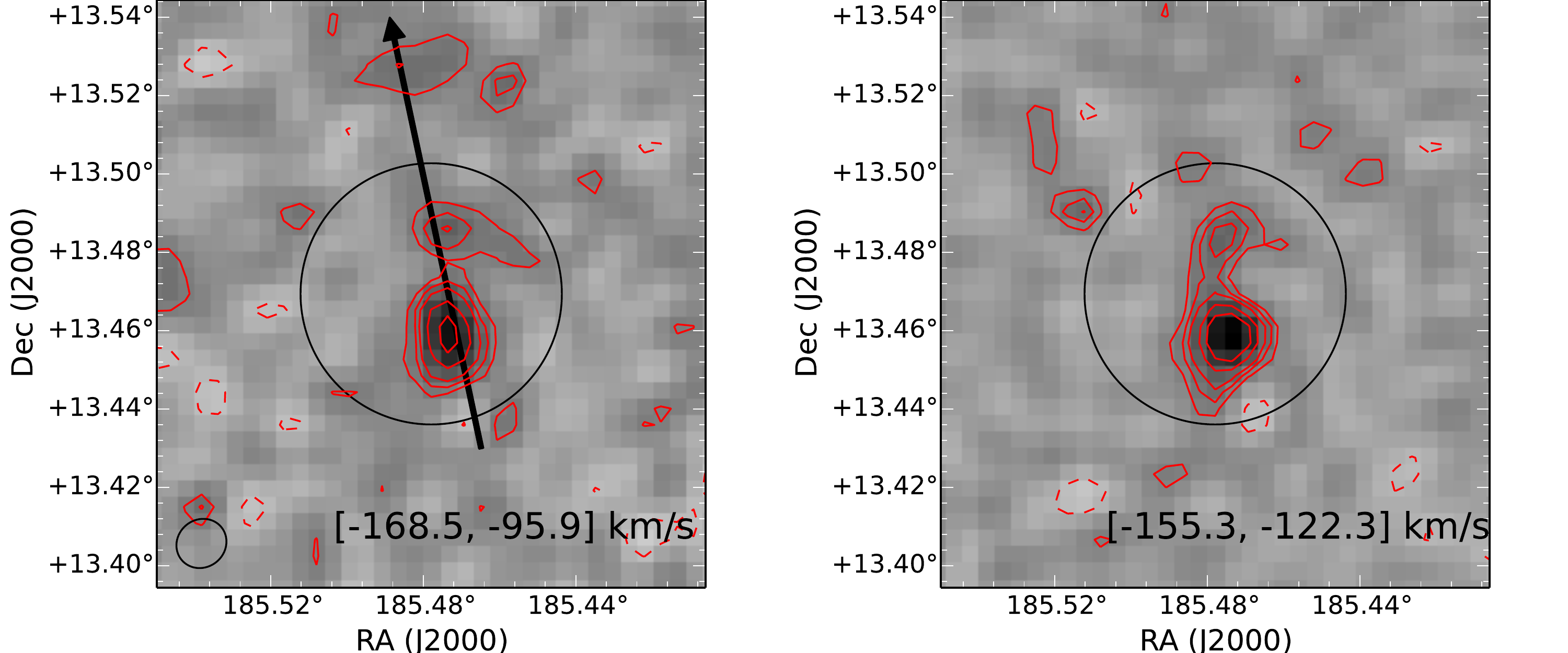}\quad
\includegraphics[width=0.46\linewidth,trim=0cm 0cm 0cm 0cm,clip=true]{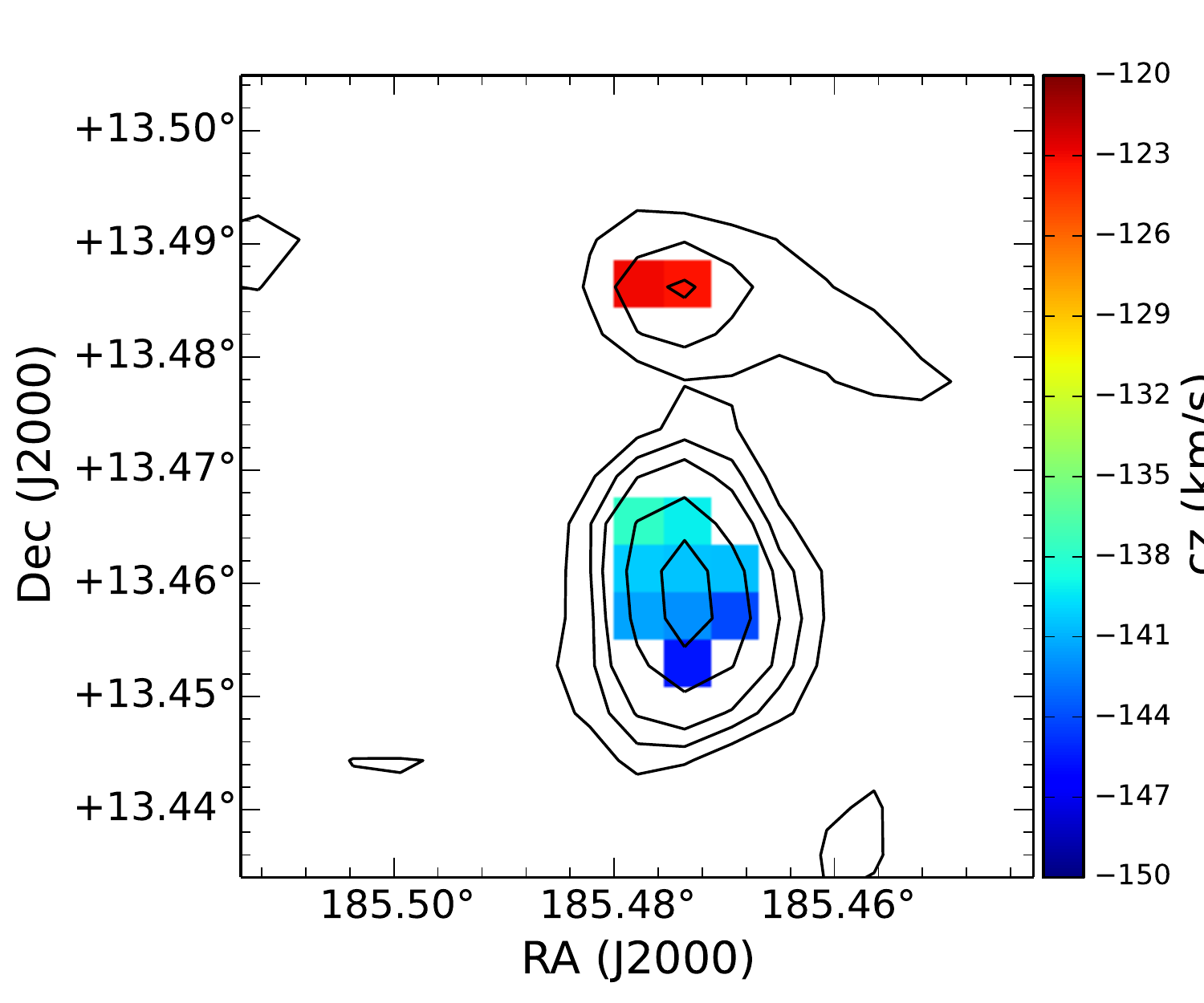}
%\quad
\quad
\includegraphics[width=0.39\linewidth,trim=0cm 0cm 1cm 1cm,clip=true]{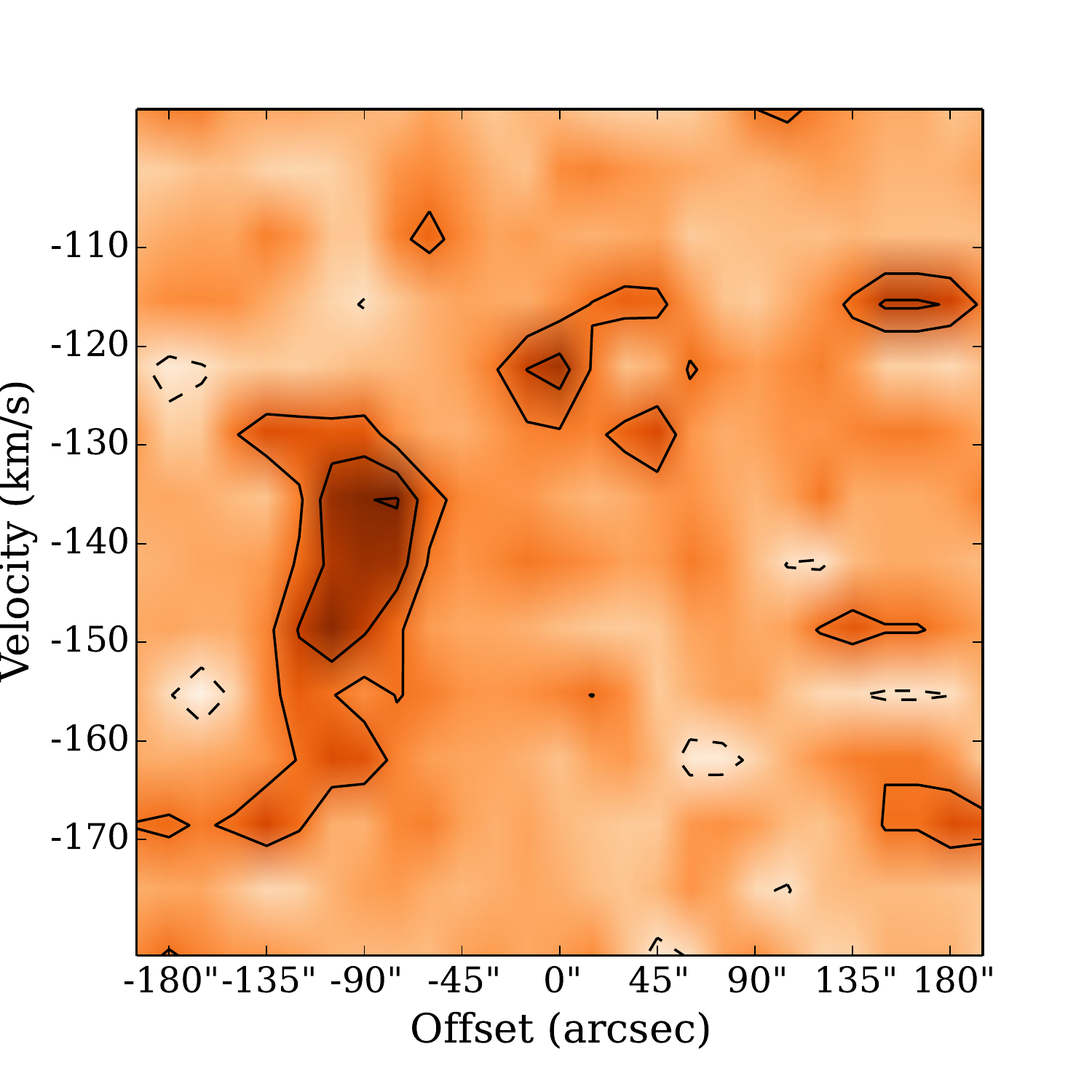}
\quad
\caption{{\it Top:} Total intensity \hi\ maps for the two channel ranges. Contours show the (-2, 2, 3, 4, 6, 8) $\times$ rms value
for each map (0.018 and 0.015 Jy \kms). 
The 4\arcmin\ AO beam is shown as the large circle for reference.  The synthesized VLA beam is shown as the small circle in the top-left figure.
 {\it Bottom Left:} \hi\ intensity contours from the full spectral range \hi\ map (upper left) overlaid
 on the velocity field from Gaussian fitting. 
  {\it Bottom right:}
Position-velocity slice along the arrow shown in the upper left with contours at (-2, 2, 4, 6) $\times$ rms (1 mJy bm$^{-1}$).
 %Constant column density contours of (2.5, 4, 5.5, 7, 8.5, 10) $\times 10^{19}$ atoms cm$^{-2}$ for the three velocity ranges overlaid on the pODI $i'$ image.
 }
\label{fig:himaps}
\end{figure*}

Imaging was done in CASA using Briggs weighting with a robust value of 0.2; the resulting
clean beam was $47.3^{\prime\prime} \times 43.8^{\prime\prime}$. Clean data cubes binned to a velocity resolution of 6.6 \kms\ (8 channels)
were produced by selecting a clean box that contained all emission from the source and cleaning to 0.5 times the
rms (1 mJy bm$^{-1}$)
over the full velocity range of emission (-168.5 to -95.9 \kms). 
%The deep cleaning
%minimizes the effects of residuals on the final flux value. 
Total \hi\ intensity (moment zero) maps were created for two different velocity ranges:
the spectral extent based on the ALFALFA spectrum (-168.5 to -95.9 \kms)
and the spectral extent based on the VLA data (-155.3 to -122.3 \kms).
The spectral extent of the VLA data was determined by spatially smoothing the data cube to a resolution
of 100\arcsec\ and selecting channels with contiguous emission.
Both \hi\ maps are shown in Figure \ref{fig:himaps} and have the same structure:
two distinct \hi\ components with evidence for low-level emission connecting them.
In addition, the full velocity range map also shows a potential northern extension of emission.

In order to include low level emission when finding the flux density,
masks were created by smoothing the moment zero maps to 100\arcsec\ resolution
and clipping at the 3-$\sigma$ level.
These masks were then applied to original moment zero maps
in order to define regions of emission used for calculating the flux density.
The recovered flux density for the full spectral range is 0.47 Jy \kms. This
is 15\% higher than
that 
found for the limited velocity range map, indicating that there is low level emission not 
clearly evident in the edge channels in the VLA data.
In addition, for the full velocity range
moment zero map, the northern extension is clearly evident in the smoothed map and
is tenuously connected to the rest of the emission at the 3-$\sigma$ level.
If this is included as part of the source, the total flux density rises to 0.68 Jy \kms.
We discuss this further in Section \ref{sec:missinghi}.

The source breaks into two distinct \hi\ components; %plus possible extended emission.
we refer to the main, central source as AGC 226067 and assign the smaller, northern source a new
identifier in the Arecibo General Catalog (AGC): 229490.
% AGC number, 229490,
%to the smaller, northern source. 
We isolate these two sources by clipping the full spectral range \hi\ map at the 2.5-$\sigma$ level. 
The individual spectra of these sources are shown in Figure \ref{fig:spec},
and the \hi\ properties are reported in Table \ref{tab:sourceprops}.
Much of the VLA emission is at low levels surrounding these two sources and 
is not directly assigned to the individual sources.

We performed a kinematic analysis of the \hi\ by fitting Gaussian functions to the spectral data 
cube using {\tt XGAUFIT} in GIPSY with the requirement that the peak amplitude be $>$5-$\sigma$ (10 mJy bm$^{-1}$). 
The resulting velocity field is shown in Figure \ref{fig:himaps}.
% for both a fit including only high significance emission and one including low significance emission. 
AGC 226067 shows evidence for rotation with an amplitude of $\sim$15 \kms. AGC 229490 is offset
by 19 \kms\ in central velocity and has narrow fitted velocity dispersions, consistent
with the velocity width found in the extracted spectrum but unusual for extragalactic systems.
A position-velocity slice bisecting AGC 226067, AGC 229490, and part of the potential northern extension
shows that the \hi\ components may be continuous in position-velocity space.

%, and the low level emission reveals tentative evidence that the two component are connected in velocity structure.
We use these data  %resolution offered by the VLA data 
to place  limits on the dynamical masses. 
%ofthe two components of this system. 
The velocity dispersion is an important contribution for these systems and we 
%We 
adopt the expression of \citetads{1996ApJS..105..269H}:
\be \label{eq:mdyn}
M_{\mathrm{dyn}} = 2.325 \times 10^5 \left( \frac{ V_{rot}^2 + 3 \sigma^2}{\mathrm{km^{2} \, s^{-2}}} \right) \left(\mathrm{\frac{r}{kpc}} \right) M_{\odot}.
\ee
For AGC 226067, we adopt a rotational velocity of 15 \kms, a velocity dispersion of 9 \kms\ based on the Gaussian fitting
and $W_{50}$ of the extracted spectrum, and a radius of 
4.1 kpc (50\arcsec\ at 17 Mpc), to find a dynamical mass of $4.5 \times 10^8$ \msun. 
For AGC 229490, we assume no rotation and adopt a velocity dispersion of 4 \kms\ and a radius of
1.6 kpc (20\arcsec\ at 17 Mpc, slightly smaller than the beam size), giving a dynamical mass estimate of $1.8 \times 10^7$ \msun.
If the velocity width or \hi\ size of AGC 229490 is larger than revealed by the low S/N data here, its dynamical mass will also be larger. Conversely, if AGC 229490 is significantly smaller than the beam of these observations, the dynamical mass will decrease.

\begin{table}
\footnotesize
\caption{Source Properties}
\label{tab:sourceprops}
\centering
\begin{tabular}{llll}
\hline
\hline
Property & AGC 226067 & AGC 229490 & AGC 229491 \\
%Source & HI Centroid & $S_{HI}$ & $W_{50}$ & $\theta_{HI}$ & $M_{HI}$\tablefootmark{a} & $M_{dyn}$\\
% & J2000 & Jy \kms & \kms & \arcsec $\times$ \arcsec & \msun & \msun \\
\hline
%$S_HI$ (Jy \kms) & 0. 28 & 0.05 & $<0.08$ \tablefootmark{a}\\
R.A. (J2000) & 12:21:53.6\tablefootmark{a}   & 12:21:53.8\tablefootmark{a}   & 12:21:55.9\tablefootmark{b} \\
Decl. (J2000) & +13:27:32\tablefootmark{a}  & +13:29:08\tablefootmark{a}  & +13:29:04\tablefootmark{b} \\
$S_{HI}$ (Jy \kms) & $0.22 \pm 0.06$ & $0.053 \pm 0.04$ & $<0.08$ \\
$cz_{\odot}$ (\kms) & -142 & -123 & ...\\
$W_{50}$ (\kms) & $25 \pm 3$   & $10 \pm 2$ & ... \\
$\sigma_{gas}$ (\kms) & $9 \pm 3$ & $4 \pm 2$ & ...\\
$a\times b$ (\hi) & 105\arcsec$\times$ 75\arcsec & 45\arcsec$\times$ 45\arcsec\tablefootmark{c} & ...\\
$N_{HI}$  (atoms cm$^{-2}$) & $8.6 \times 10^{19}$ & $4.0\times 10^{19}$ & $<2.8 \times 10^{19}$ \\ 
%\hline
$g_0$\tablefootmark{d} (mag) & $20.30 \pm 0.03$ & $\mu_g>27.6$& $22.38 \pm 0.03$\\
$(g-i)_0$\tablefootmark{d} (mag) & $0.58 \pm 0.04$ & ... & $-0.29 \pm 0.08$\\
%$\mu_{g'}$\tablefootmark{e} & 27.96 & >29.24 & 28.24\\
\mhi/$L_{g}$\tablefootmark{d} (\msun/\lsun)& $\sim$6 & $>40$\tablefootmark{e} & $< 14 $ \\
\hline
\multicolumn{4}{c}{Distance-Dependent Parameters ($D=17$ Mpc)} \\
\hline
$r_{HI}$ (kpc) & 3.7 & $<1.6$ & ...\\
\mhi\ ($10^{5}$ \msun) & $150$ & $36$ & $<50$\tablefootmark{f}\\
\mdyn\ ($10^{6}$ \msun) & $450$ & $18$ & ...\\
\mhi/\mdyn & 0.03 & 0.2 & ...\\
$M_{g}$\tablefootmark{d} (mag) & -10.85 & $>-7.2$\tablefootmark{e}& -8.77 \\
$L_{g}$\tablefootmark{d} ($10^{5}$ \lsun)& 24 & $<0.85$\tablefootmark{e} & 3.6 \\
\hline
\end{tabular}
\tablefoot{
\tablefoottext{a}{\hi\ centroid}
\tablefoottext{b}{Optical center}
\tablefoottext{c}{Unresolved by VLA beam}
\tablefoottext{d}{Corrected for Galactic extinction from \citetads{2011ApJ...737..103S}}
%\tablefoottext{e}{Average surface brightness calculated using the aperture of radius 21.34\arcsec.}
\tablefoottext{e}{Based on a 3\arcsec\ aperture and $\mu_{g} > 27.6$ mag arcsec$^{-2}$}
\tablefoottext{f}{For an unresolved source with velocity width of 26.4 \kms}
%\tablefoottext{g}{From \citetads{2009MNRAS.400.1181Z}}
%\tablefoottext{f}{Not corrected for Galactic extinction}

}
\end{table}

\subsection{WIYN/pODI data}\label{sec:podi}
\begin{figure}
\centering
\includegraphics[width=\linewidth,trim=0cm 1cm 0cm 0.25cm,clip=true]{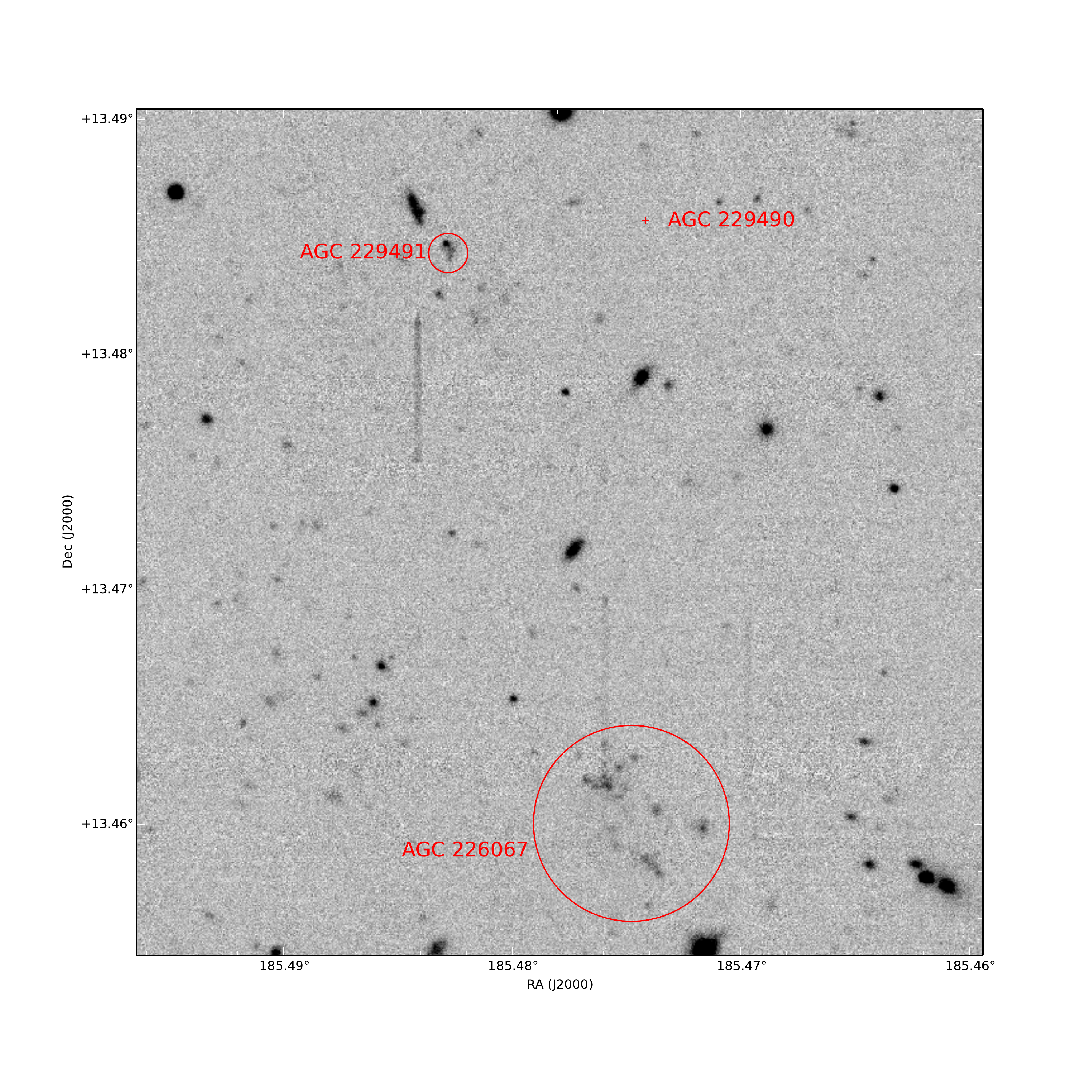}
\caption{The WIYN/pODI $g$ band image centered on the three sources of interest. The circles indicate the apertures used for photometry for AGC 226067 and AGC 229491, and the red cross the marks centroid of the \hi-only component, AGC 229490.}
\label{fig:bw}
\end{figure}

AGC 226067 was observed on 14 March 2013
 with the partially populated One Degree Imager (pODI;  $\sim 24^{\prime} \times 24^{\prime}$ field of view) on 
 the WIYN 3.5m telescope\footnote{The WIYN Observatory is a joint facility of the University of Wisconsin-Madison, Indiana University, the University of Missouri, and the National Optical Astronomy Observatory.}.
Nine 300-second exposures were obtained in a dither pattern in the SDSS $g$ and $i$ filters.
The data were reduced in the same manner as  described in \citetads{2015ApJ...801...96J}:
they were run through the  QuickReduce data reduction pipeline \citepads{2014ASPC..485..375K} using the 
%The WIYN pODI images were transferred to the 
ODI Pipeline, Portal, and Archive\footnote{The ODI Pipeline, Portal, and Archive system is a joint development project of the WIYN Consortium, Inc., in partnership with Indiana University's Pervasive Technology Institute (PTI) and with the National Optical Astronomy Observatory Science Data Management Program.} interface,
reprojected onto a common pixel scale, scaled to a common flux level, and stacked. 
Photometric calibration was done using SDSS stars in the images;
errors on the zero points were $<$0.02 mag. 
Figure \ref{fig:bw} shows the $g$ band image
centered on the region of interest.

\begin{figure*}
\centering
\includegraphics[width=0.9\linewidth,trim=0cm 0.25cm 0cm 3cm,clip=true]{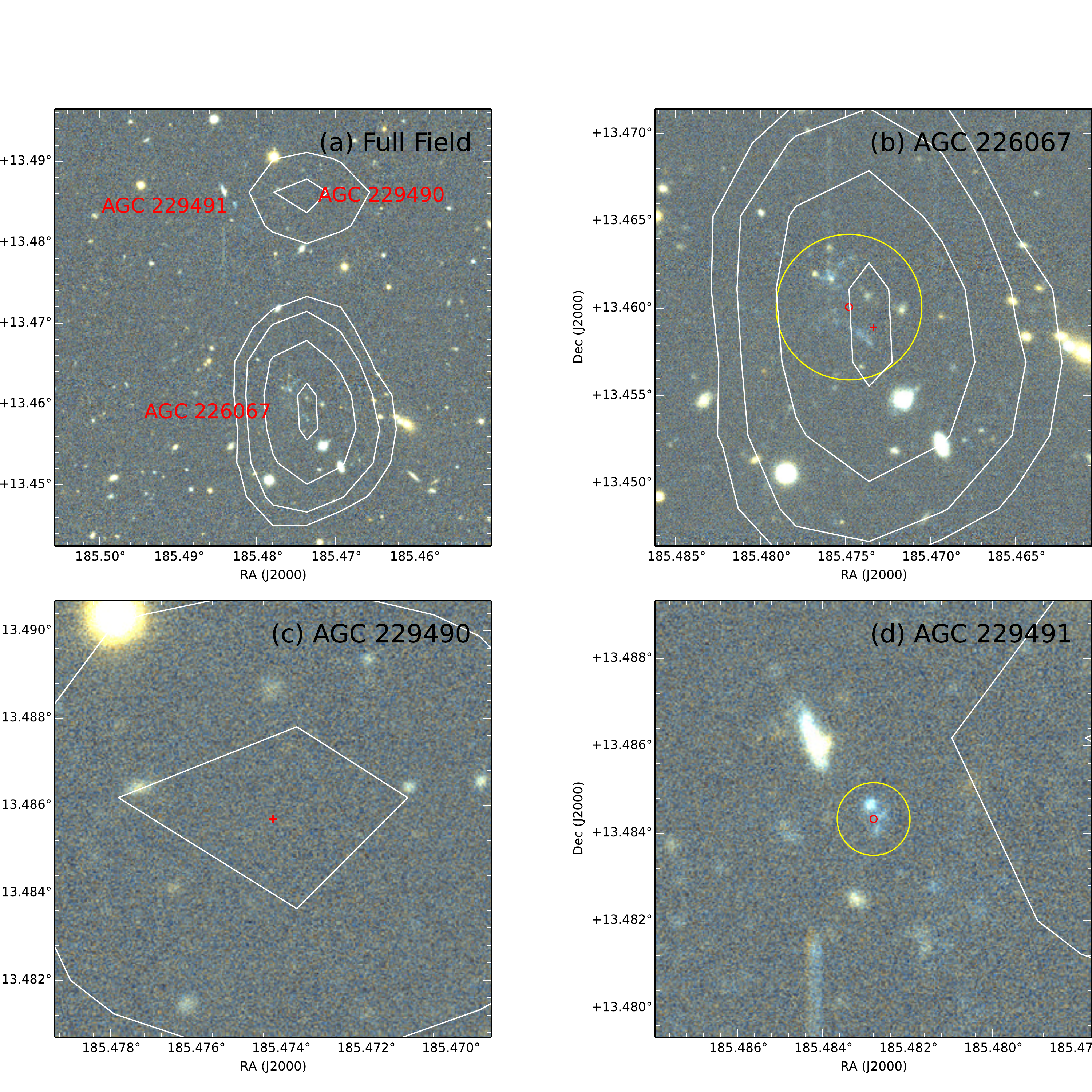}
\caption{\hi\ column density contours at (2.5, 3.5, 5.5, 8) $\times 10^{19}$ atoms cm$^{-2}$
overlaid on a WIYN/pODI color image. 
The upper left panel shows the full system, the upper right AGC 226067,
the lower left AGC 229490 (\hi-only) and the lower right AGC 229491 (optical only).
\hi\ centroids are marked with crosses and optical centroids with circles. The
yellow circles indicate the apertures used for photometry.
}
\label{fig:optical}
\end{figure*}

Figure \ref{fig:optical} shows a color image with \hi\ column density contours.
The optical counterpart 
 found by \citetalias{2015ApJ...800L..15B} (named SECCO 1)
and \citetalias{2015ApJ...806...95S} (named ALFALFA-Dw1) is 
visible and clearly associated with the main \hi\ component,
and hence
%also visible in these data.
%Given the clear association with the main \hi\ component, 
we continue to refer to this source %as AGC 226067.
by 
the
original catalog designation, AGC 226067.
After carefully masking foreground stars and background galaxies,
we measure $g_0 = 20.30 \pm 0.03 $ and $(g-i)_0 = 0.58 \pm 0.04$
 for an aperture of radius 15\arcsec, empirically determined to contain all the emission.

We independently identify the second optical counterpart noted by \citetalias{2015ApJ...806...95S}.
This source is located near the second \hi\ component (AGC 229490) but offset to the East by $\sim$30\arcsec, 
so we assign
it a unique identification number: AGC 229491.
For an aperture of radius 3\arcsec\ encompassing the blue knot of emission
that is AGC 229491, 
we find $g_0= 22.38 \pm 0.03 $ and $(g-i)_0 = -0.29 \pm 0.08$.

There is no optical source apparent  at the location of the second \hi\ peak
(AGC 229490); 
by measuring the variation in the background levels at the location of AGC 229490, we
place 3-$\sigma$ surface brightness limits of
%the limiting surface brightness for two images is 
$\mu_{g} > 27.6$ and $\mu_{i} > 26.4$ 
mag arcsec$^{-2}$.
If we assume the same optical size as for AGC 229491 (radius of 3\arcsec),
this corresponds to $g > 24.0$.

%%%%%%%%%%%%%%%%%%%%%%%%%%%%%%%%%%%%%%%%%

\section{The Sources}

\subsection{AGC 226067: A SHIELD-like dwarf galaxy}
The VLA data allow an unambiguous identification of the \hi\ in AGC 226067 with the optical counterpart
previously identified by \citetalias{2015ApJ...800L..15B} and \citetalias{2015ApJ...806...95S}.
The VLA observations reveal that  only part 
of the \hi\
emission detected by ALFALFA is associated directly with AGC 226067, and
the \hi\ mass is thus reduced 
from values based on the single-dish \hi\ emission
%resulting in a reduced \hi\ mass for AGC 226067. The \hi\ mass is reduced
by a factor of $\sim$4 
to $1.5 \times 10^7$ \msun. Correspondingly, this lowers the \mhi/$L_V$ of \citetalias{2015ApJ...800L..15B} to $\sim$5 and the \mhi/\mstar\  of 
\citetalias{2015ApJ...806...95S} to $\sim$8.
From our pODI data we find  \mhi/$L_{g} \sim$6.

In many ways, AGC 226067 is typical of low mass
gas-rich dwarfs, as exemplified by the SHIELD sample of galaxies.
The apparent optical extent in our images is 25\arcsec,
or $\sim$2.1 kpc, %and
% which translates to ~ 2.1 kpc
typical of low mass dwarf galaxies.
%peak column density of $\sim$$10^{20}$ atoms cm$^{-2}$,
The total \hi\ mass of $1.5 \times 10^7$ \msun\ and rotational amplitude of 
$\sim$15 \kms
are consistent with the properties of the SHIELD galaxies 
\citepads{2011ApJ...739L..22C,2014ApJ...785....3M}.
% , in agreement
%with what is seen in other low mass star-forming dwarf galaxies. 
However, 
 AGC 226067 %has properties similar to other low mass dwarf irregulars, but 
is distinguished %from other low mass dwarf galaxies
 by its lower surface brightness, its companions, and  extended \hi\ disk.

\subsection{AGC 229490 and AGC 229491}

The VLA and deep optical data
reveal two additional sources: an \hi-only source (AGC 229490) and
an optical-only source (AGC 229491).
Both are separated from AGC 226067 by $\sim$1.6\arcmin\ (8 kpc).
AGC 229491 was noted by \citetalias{2015ApJ...806...95S}, but
they were not able to definitively associate it with the AGC 226067 system.
 The extended \hi\ nature of this system and the close proximity to AGC 229490
(0.5\arcmin, or 2.5 kpc), strongly argue for its physical association with the AGC 226067 system.
Treating the AGC 229490 and AGC 229491 as independent sources,
 we place an
upper limit on the \hi\ mass of AGC 229491 of $5.4 \times 10^6$ \msun\ by assuming it is unresolved and has a velocity extent of 26.4 \kms\ (4 channels)\footnote{This upper limit is higher than the measured 
\hi\ mass of AGC 229490 as the velocity width of AGC 229490 is smaller than assumed here.},
and an upper limit to the luminosity of AGC 229490 of $L_{g} < 0.85 \times 10^{5}$ \lsun, assuming
the same $r=3$\arcsec\ aperture as for AGC 229491.

%These two sources are offset from each other by only 0.4\arcmin (2 kpc).
Given the close proximity of these two sources, it is plausible that they are part of the same system.
AGC 229490 could be gas that has been removed from AGC 229491, either through 
internal or external processes.
%such as star formation feed back or through interaction with the intra-cluster medium of Virgo.
Alternatively, AGC 229490 and AGC 229491 could be bright peaks of the \hi\ and stellar distribution
within the same dark matter halo.
Indeed AGC 229490 shows evidence for as extended \hi\ tail at a low (2-$\sigma$) level (see Figure
\ref{fig:himaps}).
If we  consider the two sources associated, we find a global \mhi/$L_{g}$ value of $\sim$10.

The \mhi/\mdyn\ value for AGC 229490, while uncertain due to the low resolution and S/N
of the VLA data, 
is
0.2, an unusually high value for dwarf galaxies.
This raises the question:
is this gas in a dark matter halo or free-floating \hi?
Higher resolution and deeper \hi\ data, along with deeper optical data,
are needed to fully understand AGC 229490 and AGC 229491 and their relation to AGC 226067.

\subsection{\textbf{Potential Extended Emission}}\label{sec:missinghi}
The \hi\ emission seen in the VLA data around AGC 226067 and AGC 229490 
 only accounts for
 $\sim$50-60\% of the \hi\ flux seen in the single-dish observations (Table \ref{tab:globalhi}), although
 the VLA data are shallower than the single-dish observations and the reported flux densities are
 formally consistent within the errors.
There is potentially a significant amount of emission  (0.21 Jy \kms)
contained in a northern extension.
% There is potentially a significant amount of emission (0.25 Jy \kms) contained in a northern extension.
 The VLA spectrum including this extension is shown in Figure \ref{fig:spec},
and it does a better job of matching the red edge of emission in the single-dish spectra.
 However,
this emission is low S/N and separated from the main body of emission, so
%the ALFALFA \hi\ centroid by $\gtrsim$3\arcmin\ so 
it is not clear whether this is real emission.
%and
%is consistent with being part of a background-striping data artifact\footnote{This is seen in the position-velocity
%slice in Figure \ref{fig:himaps} where emission at multiple velocities is evident at the location of the northern extension.}.
% However, this emission is quite removed 

Given the extended, interacting nature of this system it is quite
plausible that there is low surface brightness emission not
recovered by the VLA observations.
%Presumably any missing emission is low surface brightness.
To test this, we adopt the largest discrepancy 
of 0.45 Jy \kms\ (between 
the ALFALFA flux density and the VLA flux density\footnote{%This does not include the northern extension
%in the VLA data and takes  the ALFALFA flux density value from \citetalias{2013ApJ...768...77A}. 
Including the northern extension
or adopting the LBW/GBT flux density value reduces the discrepancy between the VLA and
single-dish flux density values.}).
%the amount of missing emission is
%there is a flux difference between the ALFALFA data
%and the VLA of 
%of 0.45 Jy \kms. 
%; adopting the LBW flux density value or including the 
%northern extension in the VLA data reduces this discrepancy.
Averaged over an AO beam (4\arcmin), this corresponds
to a column density of $8.6 \times 10^{18}$ atoms cm$^{-2}$, below
 the sensitivity of these VLA images (rms $= 9.5 \times 10^{18}$ atoms cm$^{-2}$).
%Presumably this emission is low surface brightness. In fact, only 70\% of the VLA flux is recovered
%when clipping directly at the 3-$\sigma$ level (what column density does this correspond to!?!).
%Producing data tapered cubes to an angular resolution of 100\arcsec\ does not (IS THIS TRUE? - CHECK!)
%recover additional flux; this implies that the low surface brightness emission exists on angular scales $<100$\arcsec.
%What do tapered data cubes show? Get a flux quickly.

%%%%%%%%%%%%%%%%%%%%%%%%%%%%%%%%%%%%%%%%%

\section{Summary and Discussion}

Our combined data reveal that the ALFALFA \hi\ source
AGC 226067 is in fact an intriguing system of at least three distinct components
plus diffuse \hi\ emission.
Given its kinematic and spatial association with the Virgo Cluster,
%and lack of a resolved stellar counterpart, 
its most likely
distance is $D=17$ Mpc.
The central source, AGC 226067, is a low mass dwarf irregular with extremely low surface
brightness, an extended \hi\ disk,  \mhi\ $= 1.5 \times 10^7$ \msun, \mhi/$L_{g} \sim$6, and \mhi/\mdyn\ $=0.03$.
The \hi-only source, AGC 229490, is possibly connected to AGC 226067 via an \hi\ bridge and
has \mhi\ $= 3.6\times 10^6$ \msun\
 and \mhi/\mdyn $\sim 0.2$.
The second optical source, AGC 229491, is extremely blue, consistent with a young
stellar population, has $L_g = 3.6\times 10^5$ \lsun, and is offset from AGC 229490 by
only 0.5\arcmin. 
%It has \mhi\ $< 5\times 10^6$ \msun\ and \mhi/$L_{g} < 14$.
The \hi\ data give a strong indication that this system
in undergoing some form of interaction.
The evidence supporting an interaction includes the putative
\hi\ bridge, a possible \hi\ tail as part of AGC 229490, and the potential
extended emission to the North.
However, the precise nature of that interaction is uncertain.
Deeper and higher resolution data, both \hi\ and optical,
are needed to fully address the origin of this system.
Here we take a preliminary look at three possibilities:
%There is clear interaction in this system with a tentative \hi\ bridge connecting 
%AGC 226067 and AGC 229490, and a kinematic continuation between the two systems.
%AGC 229491 is extremely blue, indicating it is dominated by very recent star formation.
%The two systems also represent a smooth continuation in velocity.
%We consider three possibilities:
\begin{itemize}
\item Multiple dwarf galaxy system: AGC 226067, AGC 229490 and AGC 229491 are two or three 
interacting dwarf galaxies.
%(depending
%upon the relation between AGC 229490 and AGC 229491) bound dwarf galaxies. The galaxies
%are interacting with each, resulting in the \hi\ bridge. 
One possibility is that the AGC 229490/91
system is a single dwarf galaxy that passed through AGC 226067, resulting in the offset of the
gas and stars through ram pressure.  This is
consistent with the potential \hi\ tail of AGC 229490. It is difficult to explain the regular \hi\ appearance of
AGC 226067 in this scenario, but deeper \hi\ observations may reveal an irregular \hi\ morphology.
%, undergoing
%interactions with each other.
\item Accretion onto AGC 226067: AGC 229490 is not gas in a dark matter halo
but rather is \hi\
being accreted onto AGC 226067. In this case, AGC 226067 may be undergoing triggered star formation
and only recently have developed an optical counterpart, previously being ``too shy to shine''.
However, this scenario does not explain AGC 229491.
\item Disruption of AGC 226067: AGC 229490 and AGC 229491 were originally part of AGC 226067, 
but have been
stripped from the central galaxy, either through tidal effects or ram pressure. %, presumably as it interacts with the intra-cluster medium of Virgo.
It is not clear what interaction could have led to the projected orientations of the three components.
%It may be difficult to understand how AGC 229491 ended up completely offset from the \hi\ in this scenario.
\end{itemize}

It is worth noting that several other systems with extreme \mhi/L values lie toward
the Virgo Cluster, though they are projected further from the cluster center than AGC 226067.
This includes HI1225+01,  a star-forming dwarf and a dark \hi\ cloud 
\citepads{1995AJ....109.2415C,1991AJ....101.1258S},
AGC 226178, 
a dwarf galaxy with a faint optical counterpart offset from a second \hi-poor dwarf galaxy 
\citepads{2015AJ....149...72C},
and HI1232+20,
a system of three \hi\ components of which only one has a detected low surface
brightness counterpart 
\citepads{2015ApJ...801...96J}.
%Indeed, there may still be more of these intriguing systems to be uncovered within the ALFALFA dataset
%(Leisman \etal\, in prep.).

%Note that other similar systems (multiple \hi\ components, offset optical sources, etc.) are
%seen in the Virgo detection (Coma P, AD from JC's paper, .....)
%Need better \hi\ data! 

%%%%%%%%%%%%%%%%%%%%%%%%%%%%%%%%%%%%%%%%%

\begin{acknowledgements}
We thank M. Johnson for support of the GBT observations, D. Harbeck for support of the WIYN/pODI observations, and the anonymous referee for valuable input.
The National Radio Astronomy Observatory is a facility of the National Science Foundation operated under cooperative agreement by Associated Universities, Inc.
Based on observations at Kitt Peak National Observatory, National Optical Astronomy Observatory (NOAO Prop. ID:2013A-0253; PI:Adams), which is operated by the Association of Universities for Research in Astronomy (AURA) under a cooperative agreement with the National Science Foundation.
The ALFALFA work at Cornell is supported by NSF grants AST-0607007 and AST-1107390 to R.G. and M.P.H. and by grants from the Brinson Foundation. 
K.L.R. and W.F.J. acknowledge support from NSF CAREER award AST-0847109. J.M.C. is supported by NSF grant AST-1211683.
This research made use of APLpy, an open-source plotting package for Python hosted at http://aplpy.github.com; Astropy, a community-developed core Python package for Astronomy (Astropy Collaboration, 2013); and NASA's Astrophysics Data System.

\end{acknowledgements}

%%%%%%%%%%%%%%%%%%%%%%%%%%%%%%%%%

\bibliographystyle{aa}
\bibliography{refs}

%%%%%%%%%%%%%%%%%%%%%%%%%%%%%%%%%%%%%%%%%%%

\end{document}